\documentclass[twocolumn]{aastex631}

\usepackage{natbib}
\usepackage{graphicx}
\usepackage{graphics}
\usepackage{amsmath}
\usepackage{color}
\definecolor{lblue}{rgb}{0,0.2,0.6}
\definecolor{dgreen}{rgb}{0.1,0.6,0.3}
\usepackage{enumitem}

\usepackage{threeparttable}
\usepackage{tabularx}

\usepackage{nameref}
\usepackage{varioref}
\usepackage{hyperref}
\usepackage{cleveref}
\usepackage{amsmath,amssymb,amsxtra,amsfonts}   
\usepackage{txfonts}

\newcommand       \be           {\begin{equation}}
\newcommand       \ee           {\end{equation}}

\begin{document}

\title{The UV luminosity function at $0.6<\lowercase{z}<1$ from UVCANDELS}

\correspondingauthor{Lei Sun, Xin Wang}
\email{sunl@ucas.ac.cn \quad xwang@ucas.ac.cn}

\author[0009-0004-6325-7839]{Lei Sun}
\affil{School of Astronomy and Space Science, University of Chinese Academy of Sciences, Beijing 100049, China}

\author[0000-0002-9373-3865]{Xin Wang}
\affil{School of Astronomy and Space Science, University of Chinese Academy of Sciences, Beijing 100049, China}
\affil{Institute for Frontiers in Astronomy and Astrophysics, Beijing Normal University,  Beijing 102206, China}
\affil{National Astronomical Observatories, Chinese Academy of Sciences, Beijing 100101, China}

\author[0000-0002-7064-5424]{Harry I. Teplitz}
\affil{IPAC, Mail Code 314-6, California Institute of Technology, 1200 E. California Blvd., Pasadena CA, 91125, USA}

\author[0000-0001-7166-6035]{Vihang Mehta}
\affiliation{IPAC, Mail Code 314-6, California Institute of Technology, 1200 E. California Blvd., Pasadena CA, 91125, USA}

\author[0000-0002-8630-6435]{Anahita Alavi}
\affiliation{IPAC, Mail Code 314-6, California Institute of Technology, 1200 E. California Blvd., Pasadena CA, 91125, USA}

\author[0000-0002-9946-4731]{Marc Rafelski}
\affiliation{Space Telescope Science Institute, Baltimore, MD 21218, USA}
\affiliation{Department of Physics and Astronomy, Johns Hopkins University, Baltimore, MD 21218, USA}

\author[0000-0001-8156-6281]{Rogier A. Windhorst} 
\affiliation{School of Earth and Space Exploration, Arizona State University, Tempe, AZ 85287, USA}

\author[0000-0002-9136-8876]{Claudia Scarlata}
\affiliation{Minnesota Institute of Astrophysics and School of Physics and Astronomy, University of Minnesota, Minneapolis, MN 55455, USA}

\author[0000-0003-2098-9568]{Jonathan P. Gardner}
\affiliation{Astrophysics Science Division, NASA Goddard Space Flight Center, Greenbelt, MD 20771, USA}

\author[0000-0002-0648-1699]{Brent M. Smith}
\affil{School of Earth and Space Exploration, Arizona State University, Tempe, AZ 85287, USA}

\author[0000-0003-3759-8707]{Ben Sunnquist}
\affiliation{Space Telescope Science Institute, Baltimore, MD 21218, USA}

\author[0000-0002-0604-654X]{Laura Prichard}
\affiliation{Space Telescope Science Institute, Baltimore, MD 21218, USA}

\author[0000-0001-8551-071X]{Yingjie Cheng}
\affiliation{Department of Astronomy, University of Massachusetts, Amherst, 01003 MA, USA}

\author[0000-0001-9440-8872]{Norman Grogin}
\affiliation{Space Telescope Science Institute, Baltimore, MD 21218, USA}

\author[0000-0001-6145-5090]{Nimish P. Hathi}
\affiliation{Space Telescope Science Institute, Baltimore, MD 21218, USA}

\author[0000-0001-8587-218X]{Matthew Hayes}
\affiliation{Stockholm University, Department of Astronomy and Oskar Klein Centre for Cosmoparticle Physics, AlbaNova University Centre, SE-10691, Stockholm, Sweden}

\author[0000-0002-6610-2048]{Anton M. Koekemoer}
\affiliation{Space Telescope Science Institute, Baltimore, MD 21218, USA}

\author[0000-0001-5846-4404]{Bahram Mobasher}
\affiliation{Department of Physics and Astronomy, University of California, Riverside, Riverside, CA 92521, USA}

\author[0000-0001-5294-8002]{Kalina V. Nedkova}
\affiliation{Department of Physics and Astronomy, Johns Hopkins University, Baltimore, MD 21218, USA}

\author[0000-0002-8190-7573]{Robert O'Connell}
\affiliation{Department of Astronomy, University of Virginia, Charlottesville, VA 22904}

\author[0000-0002-4271-0364]{Brant Robertson}
\affiliation{Department of Astronomy and Astrophysics, University of California, Santa Cruz, Santa Cruz, CA 95064, USA}

\author[0000-0003-0749-4667]{Sina Taamoli}
\affiliation{Department of Physics and Astronomy, University of California, Riverside, Riverside, CA 92521, USA}

\author[0000-0003-3466-035X]{L. Y. Aaron Yung}
\affiliation{Astrophysics Science Division, NASA Goddard Space Flight Center, Greenbelt, MD 20771, USA}


\author[0000-0003-2680-005X]{Gabriel Brammer}
\affiliation{Cosmic Dawn Center (DAWN), Denmark}
\affiliation{Niels Bohr Institute, University of Copenhagen, Jagtvej 128, DK-2200 Copenhagen N, Denmark}

\author[0000-0001-6482-3020]{James Colbert}
\affiliation{IPAC, Mail Code 314-6, California Institute of Technology, 1200 E. California Blvd., Pasadena CA, 91125, USA}

\author[0000-0003-1949-7638]{Christopher Conselice}
\affiliation{School of Physics and Astronomy, The University of Nottingham, University Park, Nottingham NG7 2RD, UK}

\author[0000-0003-1530-8713]{Eric Gawiser}
\affiliation{Department of Physics and Astronomy, Rutgers, The State University of New Jersey, Piscataway, NJ 08854, USA}

\author[0000-0003-2775-2002]{Yicheng Guo}
\affiliation{Department of Physics and Astronomy, University of Missouri, Columbia, MO 65211, USA}

\author[0000-0003-1268-5230]{Rolf A. Jansen}
\affiliation{School of Earth and Space Exploration, Arizona State University, Tempe, AZ 85287, USA}

\author[0000-0001-7673-2257]{Zhiyuan Ji}
\affiliation{Department of Astronomy, University of Massachusetts, Amherst, MA 01003, USA}

\author[0000-0003-1581-7825]{Ray A. Lucas}
\affiliation{Space Telescope Science Institute, Baltimore, MD 21218, USA}

\author[0000-0001-7016-5220]{Michael Rutkowski}
\affiliation{Department of Physics and Astronomy, Minnesota State University Mankato, Mankato, MN 56001, USA}

\author[0000-0002-4935-9511]{Brian Siana}
\affiliation{Department of Physics and Astronomy, University of California, Riverside, Riverside, CA 92521, USA}

\author[0000-0002-5057-135X]{Eros Vanzella}
\affiliation{INAF -- OAS, Osservatorio di Astrofisica e Scienza dello Spazio di Bologna, via Gobetti 93/3, I-40129 Bologna, Italy}

\author[0000-0003-4439-6003]{Teresa Ashcraft}
\affiliation{School of Earth and Space Exploration, Arizona State University, Tempe, AZ 85287, USA}
\author[0000-0002-9921-9218]{Micaela Bagley}
\affiliation{Department of Astronomy, The University of Texas at Austin, Austin, TX 78712, USA}
\author[0000-0003-0556-2929]{Ivano Baronchelli}
\affiliation{INAF-Osservatorio Astronomico di Padova, Vicolo dell'Osservatorio 5, I-35122, Padova, Italy}
\affiliation{Dipartimento di Fisica e Astronomia, Università di Padova, vicolo Osservatorio, 3, I-35122 Padova, Italy}
\author[0000-0001-6813-875X]{Guillermo Barro}
\affiliation{Department of Physics, University of the Pacific, Stockton, CA 95211, USA}
\author[0000-0003-2102-3933]{Alex Blanche}
\affiliation{School of Earth and Space Exploration, Arizona State University, Tempe, AZ 85287, USA}
\author[0000-0002-7767-5044]{Adam Broussard}
\affiliation{Department of Physics and Astronomy, Rutgers, The State University of New Jersey, Piscataway, NJ 08854, USA}
\author[0000-0001-6650-2853]{Timothy Carleton}
\affiliation{School of Earth and Space Exploration, Arizona State University, Tempe, AZ 85287, USA}
\author[0000-0003-3691-937X]{Nima Chartab}
\affiliation{Observatories of the Carnegie Institution of Washington, Pasadena, CA 91101, US}
\author[0000-0001-9560-9174]{Alex Codoreanu}
\affiliation{Centre for Astrophysics and Supercomputing, Swinburne University of Technology, Hawthorn VIC 3122, Australia}
\author[0000-0003-3329-1337]{Seth Cohen} 
\affiliation{School of Earth and Space Exploration, Arizona State University, Tempe, AZ 85287, USA}

\author[0000-0002-7928-416X]{Y. Sophia Dai}
\affiliation{National Astronomical Observatories, Chinese Academy of Sciences, Beijing 100101, China}
\author[0000-0003-4919-9017]{Behnam Darvish}
\affiliation{Department of Physics and Astronomy, University of California, Riverside, Riverside, CA 92521, USA}
\author[0000-0003-2842-9434]{Romeel Dav\'{e}}
\affiliation{Institute for Astronomy, University of Edinburgh, Edinburgh, EH9 3HJ, UK}
\author[0000-0001-9022-665X]{Laura DeGroot}
\affiliation{College of Wooster, Wooster, OH 44691, USA}
\author{Duilia De Mello}
\affiliation{Department of Physics, The Catholic University of America, Washington, DC 20064, USA}
\author[0000-0001-5414-5131]{Mark Dickinson}
\affiliation{NSF's NOIRLab, Tucson, AZ 85719, USA}
\author[0000-0003-2047-1689]{Najmeh Emami}
\affiliation{Department of Physics and Astronomy, University of California, Riverside, Riverside, CA 92521, USA}
\author[0000-0001-7113-2738]{Henry Ferguson}
\affiliation{Space Telescope Science Institute, Baltimore, MD 21218, USA}
\author[0000-0002-8919-079X]{Leonardo Ferreira}
\affiliation{Centre for Astronomy and Particle Physics, School of Physics and Astronomy, University of Nottingham, NG7 2RD, UK}
\author[0000-0003-0792-5877]{Keely Finkelstein}
\affiliation{Department of Astronomy, The University of Texas at Austin, Austin, TX 78712, USA}
\author[0000-0001-8519-1130]{Steven Finkelstein}
\affiliation{Department of Astronomy, The University of Texas at Austin, Austin, TX 78712, USA}

\author[0000-0002-7732-9205]{Timothy Gburek}
\affiliation{Department of Physics and Astronomy, University of California, Riverside, Riverside, CA 92521, USA}
\author[0000-0002-7831-8751]{Mauro Giavalisco}
\affiliation{Department of Astronomy, University of Massachusetts, Amherst, MA 01003, USA}
\author[0000-0002-5688-0663]{Andrea Grazian}
\affiliation{INAF-Osservatorio Astronomico di Padova, Vicolo dell'Osservatorio 5, I-35122, Padova, Italy}
\author[0000-0001-6842-2371]{Caryl Gronwall}
\affiliation{Department of Astronomy \& Astrophysics, The Pennsylvania State University, University Park, PA 16802, USA}
\affiliation{Institute for Gravitation and the Cosmos, The Pennsylvania State University, University Park, PA 16802, USA}
\author[0000-0003-2226-5395]{Shoubaneh Hemmati}
\affiliation{IPAC, Mail Code 314-6, California Institute of Technology, 1200 E. California Blvd., Pasadena CA, 91125, USA}
\author[0000-0002-5924-0629]{Justin Howell}
\affiliation{IPAC, Mail Code 314-6, California Institute of Technology, 1200 E. California Blvd., Pasadena CA, 91125, USA}
\author[0000-0001-9298-3523]{Kartheik Iyer}
\affiliation{Dunlap Institute for Astronomy \& Astrophysics, University of Toronto, 50 St George Street, Toronto, ON M5S 3H4, CA}
\author[0000-0002-5601-575X]{Sugata Kaviraj}
\affiliation{Centre for Astrophysics Research, Department of Physics, Astronomy and Mathematics, University of Hertfordshire, Hatfield, AL10 9AB, UK}
\author[0000-0002-8816-5146]{Peter Kurczynski}
\affiliation{Astrophysics Science Division, NASA Goddard Space Flight Center, Greenbelt, MD 20771, USA}
\author[0009-0000-1797-0300]{Ilin Lazar}
\affiliation{Department of Galaxies and Cosmology, Max Planck Institute for Astronomy, K\"{o}nigstuhl 17, 69117 Heidelberg}
\author[0000-0001-6529-8416]{John MacKenty}
\affiliation{Space Telescope Science Institute, Baltimore, MD 21218, USA}
\author[0000-0002-6016-300X]{Kameswara Bharadwaj Mantha}
\affiliation{Minnesota Institute of Astrophysics and School of Physics and Astronomy, University of Minnesota, Minneapolis, MN 55455, USA}
\author[0000-0002-6632-4046]{Alec Martin}
\affiliation{Department of Physics and Astronomy, University of Missouri, Columbia, MO 65211, USA}
\author[0000-0003-2939-8668]{Garreth Martin}
\affiliation{Korea Astronomy and Space Science Institute, Yuseong-gu, Daejeon 34055, Korea}
\affiliation{Steward Observatory, University of Arizona, Tucson, AZ 85719, USA}
\author[0000-0002-5506-3880]{Tyler McCabe}
\affiliation{School of Earth and Space Exploration, Arizona State University, Tempe, AZ 85287, USA}

\author[0000-0002-8085-7578]{Charlotte Olsen}
\affiliation{Department of Physics and Astronomy, Rutgers, The State University of New Jersey, Piscataway, NJ 08854, USA}
\author[0000-0001-9665-3003]{Lillian Otteson}
\affiliation{School of Earth and Space Exploration, Arizona State University, Tempe, AZ 85287, USA}
\author[0000-0002-5269-6527]{Swara Ravindranath}
\affiliation{Space Telescope Science Institute, Baltimore, MD 21218, USA}
\author[0000-0002-9961-2984]{Caleb Redshaw}
\affiliation{School of Earth and Space Exploration, Arizona State University, Tempe, AZ 85287, USA}
\author[0000-0002-0364-1159]{Zahra Sattari}
\affiliation{Department of Physics and Astronomy, University of California, Riverside, Riverside, CA 92521, USA}
\author[0000-0002-2390-0584]{Emmaris Soto}
\affiliation{Computational Physics, Inc., Springfield, VA 22151, USA}
\author[0000-0002-7830-363X]{Bonnabelle Zabelle}
\affiliation{Minnesota Institute of Astrophysics and School of Physics and Astronomy, University of Minnesota, Minneapolis, MN 55455, USA}
\author{the UVCANDELS team}

\begin{abstract}

UVCANDELS is a HST Cycle-26 Treasury Program awarded 164 orbits of primary ultraviolet (UV) F275W imaging and coordinated parallel optical F435W imaging in four CANDELS fields: GOODS-N, GOODS-S, EGS, and COSMOS, covering a total area of $\sim426$ arcmin$^2$. This is $\sim2.7$ times larger than the area covered by previous deep-field space UV data combined, reaching a depth of about 27 and 28 ABmag ($5\sigma$ in $0."2$ apertures) for F275W and F435W, respectively. Along with the new photometric catalogs, we present an analysis of the rest-frame UV luminosity function (LF), relying on our UV-optimized aperture photometry method yielding a factor of $1.5\times$ increase than the H-isophot aperture photometry in the signal-to-noise ratios of galaxies in our F275W imaging. Using well tested photometric redshift measurements we identify 5810 galaxies at redshifts $0.6<z<1$, down to an absolute magnitude of $M_\text{UV} = -14.2$. In order to minimize the effect of uncertainties in estimating the completeness function, especially at the faint-end, we restrict our analysis to sources above $30\%$ completeness, which provides a final sample of 4726 galaxies at $-21.5<M_\text{UV}<-15.5$. We performed a maximum likelihood estimate to derive the best-fit parameters of the UV LF. We report a best-fit faint-end slope of $\alpha = -1.359^{+0.041}_{-0.041}$ at $z \sim 0.8$. Creating sub-samples at $z\sim0.7$ and $z\sim0.9$, we observe a possible evolution of $\alpha$ with redshift. The unobscured UV luminosity density at $M_\text{UV}<-10$ is derived as $\rho_\text{UV}=1.339^{+0.027}_{-0.030}\ (\times10^{26} \text{ergs/s/Hz/Mpc}^3)$ using our best-fit LF parameters. The new F275W and F435 photometric catalogs from UVCANDELS have been made publicly available on the Barbara A. Mikulski Archive for Space Telescopes (MAST).

\end{abstract}


\section{Introduction}
\label{sec:intro}

The luminosity function (LF) of galaxies is one of the key probes of galaxy formation and evolution, as the shape of the LF is mainly determined by the underlying halo mass function and the mechanisms that regulate star formation in galaxies, such as gas cooling and feedback processes \citep{Rees77, White78, Benson03}. The LF is also an important tool to evaluate the contribution of galaxies with different luminosities to the cosmic light budget as a function of redshift. The rest-frame ultraviolet (UV) light is a direct tracer of recent star formation in galaxies; therefore, the rest-frame UV LF can be used to determine the volume-averaged cosmic star formation rate. Moreover, the UV LF is directly measurable up to very high redshifts, $z\sim10$ and beyond, which makes it a reliable technique to investigate star formation and mass build up in galaxies out to very early epochs. 

Recently, measurements of the rest-frame UV LFs from 
intermediate redshift $z\sim2$ up to very high $z\sim16$ have increased dramatically as new facilities such as WFC3 on HST and the most up-to-date JWST bring forth data products \citep{Hathi10, Finkelstein15, Alavi16, Bouwens16, Mcleod16, Stefanon2017, mehtaUVUDFUVLuminosity2017, Ono18, Ishigaki18, Kawamata18, Viironen18, Pello18, Yue18, Bhatawdekar19, Khusanova20, Rojas20, Bowler20, Adams20, Ito20, Zhang21, Bouwens21, Harikane22, Bagley22, Bouwens22, Finkelstein22a, Finkelstein22b, Finkelstein23, Leethochawalit23, Donnan23, Harikane23, Adams23a, Adams23b, Mcleod23, Leung23, Perez23, Varadaraj23, Harikane24}.
As studies accumulate, the low luminosity galaxies have attracted a large amount of 
attention. At $z\sim6-10$, faint galaxies are expected to play a dominant role in the reionization of the universe \citep[e.g.,][]{Bouwens12, Robertson15, Yung20b, Bouwens22}. At intermediate redshifts, they are also crucial in investigating feedback due to star formation and reionization \citep[e.g.,][]{Weinmann12, Yung20a, Yung20b}. Employing HST data and taking advantage of lensing magnification by the Hubble Frontier Field clusters to probe the faint end of the UV LFs, \citet{Bouwens22} unveil that the faint end slope $\alpha$ gets smoothly steeper from $-1.53\pm0.03$ at $z=2$ to $-2.28\pm0.10$ at $z=9$. Making use of the JWST early release observations data, \citet{Harikane23} push the measuring of the UV LFs to very high redshift $z\sim9-16$ at the pre-reionization stage, and find agreement with other HST and JWST studies.

Fewer studies are available of the rest-UV LFs at relatively low redshift $z<2$.
\citet{Arnouts05} use the FUV and NUV data from the GLAEX VVDS observations to measure the UV LFs at $0.2<z<1.2$. \citet{Oesch10} target the redshift range $0.5<z<1$ using the HST WFC3 Early Release Science (ERS) data in the GOODS-S field. \citet{Cucciati12} use data from the VVDS to extend the redshift range to $0.05-4.5$. \citet{Moutard20} explore the range of $z\sim0.2-3$ using data from the CFHT Large Area U-band Deep Survey (CLAUDS) and the HyperSuprime-Cam Subaru Strategic Program (HSC-SSP). \citet{Page21}, \citet{Sharma22a} and \citet{Sharma22b} focus on $0.6<z<1.2$ using data from the XMM-OM observations. Very recently, \citet{Bhattacharya23} present their results at $0.01<z<0.79$ from both the Astrosat and the HST observations. 

At these relatively low redshifts, the survey flux limit are still being improved, approaching the fainter end of the UV LFs, e.g., \citet{Weisz14} reconstruct the UV LFs down to very faint magnitudes $-14<M_{UV}<-1.5$ using the local group fossil records. Lensing magnification by foreground systems can also further improve our ability to probe the faint end of LFs at relatively low redshift. Employing 3 lensing galaxy clusters, \citet{Alavi16} measure the UV LF and its evolution during the peak epoch of cosmic star formation at $1<z<3$ down to $M_{UV}=-12.5$. 
Therefore, we can obtain better constraints at these redshifts on the faint end slope $\alpha$ and its evolution with redshift, helping to establish the cosmic star formation history and to suggest the role of the high-z analogs of these low luminosity galaxies in the cosmic reionization.

The Ultraviolet Imaging of the Cosmic Assembly Near-infrared Deep Extragalactic Legacy Survey Fields (UVCANDELS; GO-15647, PI: Teplitz; \citealt{wangLymanContinuumEscape2023}) survey is a HST Cycle-26 Treasury Program awarded 164 orbits of primary ultraviolet (UV) WFC3 F275W imaging and coordinated parallel optical ACS F435W imaging in four CANDELS fields \citep{groginCANDELSCOSMICASSEMBLY2011, koekemoerCANDELSCOSMICASSEMBLY2011}: GOODS-N, GOODS-S, EGS, and COSMOS, covering a total area of $\sim426$ arcmin$^2$. This is $\sim2.7$ times larger than the area covered by previous deep-field space UV data combined. UVCANDELS takes F275W exposures at a
uniform 3-orbit depth and F435W exposures at slightly varying depth resulting from the roll angle constraints and the
overlap from the increased FoV of the ACS camera,  reaching a depth of 27 and 28 ABmag ($5\sigma$ in $0.2$ arcsec apertures) for F275W and F435W, respectively. Relying on our new photometric catalogs and accurate photometric redshift estimates (Mehta et al. in prep.), UVCANDELS is capable of constraining the key parameters of the UV LFs with a high precision at relatively low redshift, $z\lesssim2$. The wide area coverage (over multiple fields) of UVCANDELS helps drive down the cosmic variance and thus limit the uncertainty at the bright end of the UV LF, and its competitive imaging depth, meanwhile, allows for a robust measurement of the faint end slope of the UV LF at the most intense epoch in star formation history of the Universe.

In this paper, we first present the photometric catalogs of UVCANDELS in F275W and F435W bands. We then perform an analysis of the rest-frame UV LF in the redshift range $0.6<z<1$ using the UVCANDELS F275W imaging data and provide constraints on the UV LF parameters. The structure of this paper is as follows: in Section~\ref{sec:photom} we describe our photometry methods and present the final photometry catalogs obtained from our UVCANDELS observations, Section~\ref{sec:select} describes our sample selection, in Section~\ref{sec:comp} we present our completeness analysis, we outline the procedures for deriving the best-fit UV LF and present the results in Section~\ref{sec:lf}, this is followed by a discussion of our results in Section~\ref{sec:discussion}. Our conclusions are summarized in Section~\ref{sec:summary}.

Throughout this paper, we adopt cosmological parameters from  \citet{planck18}: $\Omega_m=0.315$, $\Omega_\lambda = 0.685$ and $H_0=67.4$ km s$^{-1}$ Mpc$^{-1}$ and all magnitudes used are AB magnitudes \citep{Oke83}.

\section{Photometry and catalog creation}
\label{sec:photom}

The observation and reduction of our UVCANDELS imaging data are described in detail in \citet{wangLymanContinuumEscape2023}. Briefly, individual exposures in the F275W and F435W filters are first pre-processed to account for the effects of charge transfer inefficiency, scattered light from the earth limb, and cosmic rays, etc.. Then we coadd the flux-calibrated flat-fielded individual exposures using the AstroDrizzle software \citep{Gonzaga.2012}. The resulting UVCANDELS F275W and F435W coadded mosaics\footnote{publicly available at \url{https://archive.stsci.edu/hlsp/uvcandels} with\dataset[DOI: 10.17909/8s31-f778]{https://doi.org/10.17909/8s31-f778}} are astrometrically aligned to the world coordinate system used by the CANDELS images. 

Here we describe our aperture-matched point spread function (PSF) corrected photometry methodology employed to create the catalogs. 
The CANDELS multi-wavelength photometric catalogs for the GOODS-N, GOODS-S, COSMOS and EGS fields are presented in \citet{barroCANDELSSHARDSMultiwavelength2019}, \citet{guoCANDELSMULTIWAVELENGTHCATALOGS2013}, \citet{nayyeriCANDELSMULTIWAVELENGTHCATALOGS2017}, and \citet{stefanonCANDELSMultiwavelengthCatalogs2017}, respectively.
These catalogs include the photometric measurements of objects in the optical to mid-infrared wavelengths.
Since the primary goal of the UVCANDELS program is to complete the UV and blue-optical coverage of these four premier extragalactic legacy fields, we first follow the standard approach utilized by the CANDELS team.
We run \textsc{SExtractor} v2.8.6 \citep{Bertin.1996} in dual-image mode with the near-infrared F160W coadded mosaics \citep{groginCANDELSCOSMICASSEMBLY2011,koekemoerCANDELSCOSMICASSEMBLY2011} as the detection images.
For the measurement images, we use the PSF-matched F275W and F435W images, produced from the original science images, convolved with PSF homogenization kernels, to bring their PSF's full width half maximum (FWHM) to match that of the F160W (H-band) PSFs.
Two sets of \textsc{SExtractor} detection parameters \citep[see e.g., Table~3 in][]{barroCANDELSSHARDSMultiwavelength2019} are adopted to combine the ``hot'' plus ``cold'' source detection strategy. The ``hot'' mode is used to push the detection of faint sources to the limiting depth of the imaging mosaics and at the same time recover large/bright objects without excessive deblending using the ``cold'' mode.
An aperture correction is then applied to convert the F275W/F435W isophotal magnitudes to total magnitudes using, i.e.,
\begin{equation}
    {\rm mag}_{\rm tot}^{\rm candels} = {\rm mag}_{\rm iso}^{\circledast} + (H_{\rm auto} - H_{\rm iso}).
\end{equation}
We estimate magnitude uncertainties through proper error propagation. 
Hereafter, we refer to the F275W/F435W total magnitudes measured in this fashion (${\rm mag}_{\rm tot}^{\rm candels}$) as the results from the H-isophot aperture photometry. 
In total, we detect 3106, 1221, 3040, 3809 sources with a signal-to-noise ratio (SNR) threshold of SNR$\geq$3 in F275W in the GOODS-N, GOODS-S, COSMOS and EGS fields, respectively, using the H-isophot aperture photometry. At the 5-$\sigma$ detection level, we measure 1610, 595, 1491, 1786 objects in these fields, respectively.
In the COSMOS and EGS fields, where the new B-band imaging was taken, we obtain 12994 (8756) and 16016 (10838) sources at 3(5)-$\sigma$ significance, respectively.

{
\tabletypesize{\footnotesize}
\tabcolsep=2pt
\begin{deluxetable}{lllll}
    \label{table:sex_param}
    \tablecolumns{5}
    \tablewidth{0pt}
    \tablecaption{Values of key SExtractor parameters set in our improved aperture photometry method optimized for F275W and F435W imaging from UVCANDELS observations.}
\tablehead{
    Parameter &  GOODS-N &  GOODS-S &  COSMOS  & EGS
    }
\startdata
    \texttt{DETECT\_MINAREA}    & 5  & 5  & 5  & 5   \\
    \texttt{DETECT\_THRESH}     & 1.4  & 1.7  & 1.0  & 1.3  \\
    \texttt{ANALYSIS\_THRESH}   & 1.4  & 1.7  & 1.0  & 1.3  \\
    \texttt{FILTER\_NAME}       & gauss\_4.0 & gauss\_4.0 & gauss\_4.0 & gauss\_4.0  \\
    \texttt{DEBLEND\_NTHRESH }  & 32   & 32   & 32   & 32    \\
    \texttt{DEBLEND\_MINCONT}   & 1e-5 & 1e-5 & 1e-5 & 1e-5  \\
    \texttt{BACK\_SIZE}         & 128  & 128  & 128  & 128   \\
    \texttt{BACK\_FILTERSIZE}   & 5  & 5  & 5  & 5   \\
    \texttt{BACKPHOTO\_THICK}   & 48 & 48 & 48 & 48
\enddata
    \tablecomments{We set the values of the parameters \texttt{DETECT\_THRESH} and \texttt{ANALYSIS\_THRESH} taking into account the varying depth of the pre-existing F606W imaging in the four UVCANDELS fields. The exposure time for the F606W imaging mosaics in GOODS-N (wide region), GOODS-S (wide region), COSMOS, and EGS is 5600, 8600, 3300, and 5700 seconds, respectively.}
\end{deluxetable}
}

\begin{figure*}
    \centering
    \includegraphics[width=.8\textwidth,trim=0cm 4.8cm 0cm 4.7cm,clip]{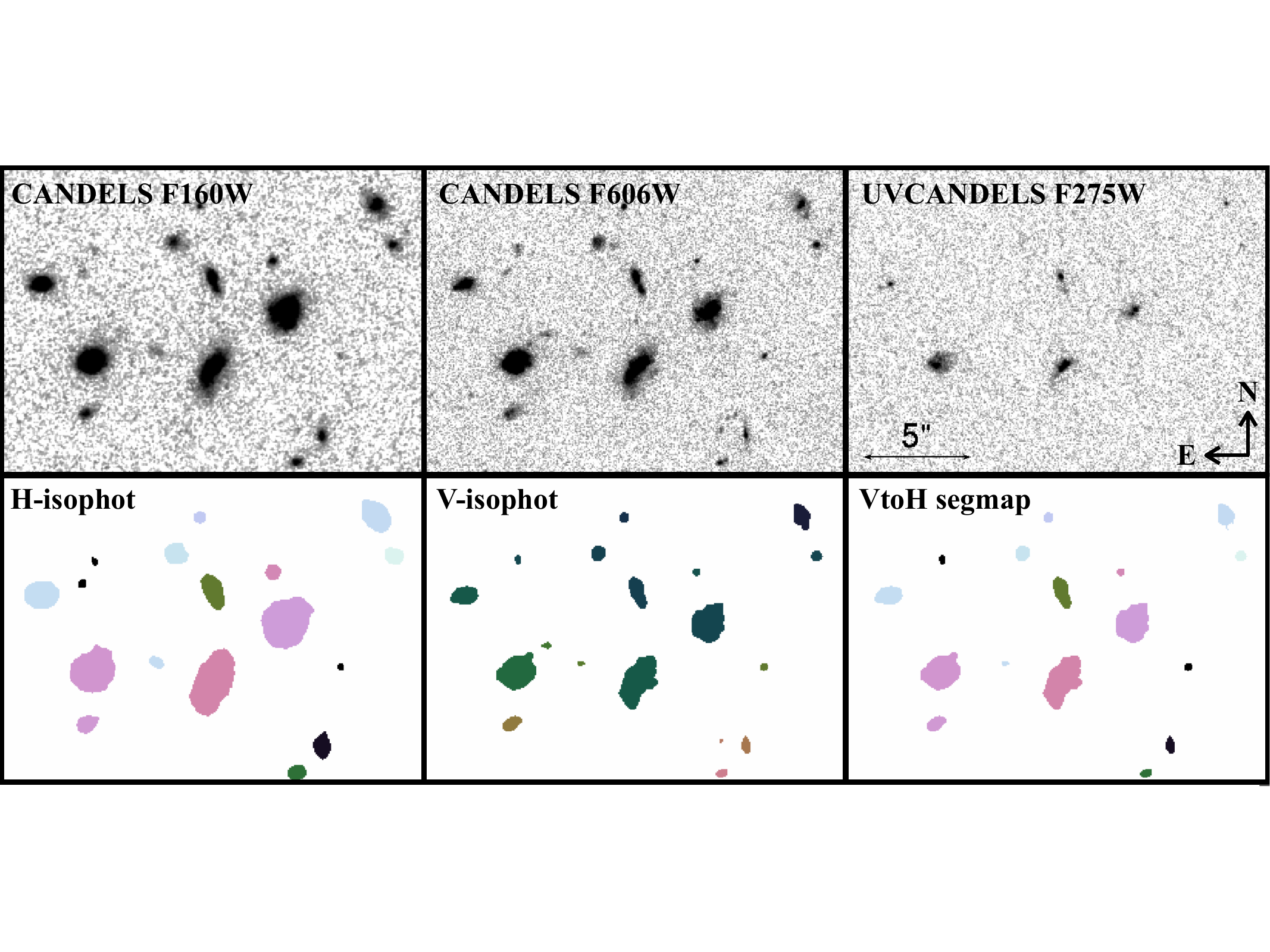}
    \caption{An illustration of our UV-optimized aperture photometry method utilized in this work. Top from left to right, we show example image stamps cut from the F160W, F606W and F275W original science mosaics.
    The left and middle panels on the bottom row show the object isophotes measured in F160W (H-band) and F606W (V-band), respectively, both color-coded in object IDs.
    The object V-band isophotes are clearly more appropriate for photometry of UV fluxes.
    As a result, we assign object IDs defined in F160W to segmentation regions obtained in F606W to get the ``VtoH'' segmentation map shown in the lower right panel, used for our V-isophot aperture photometry (see Eq.~2). See also \citet{rafelskiUvudfUltravioletNearinfrared2015} for specific segmentation algorithm and Tab.~\ref{table:sex_param}  for values of parameters used for segmentation.
    }
    \label{fig:VtoH_illustrate}
\end{figure*}

However, several downsides exist for this H-isophot aperture photometry, in particular for UV images. First and foremost, matching to the H-band PSF reduces image resolution: all measurements are conducted on convolved data (i.e. ${\rm mag}_{\rm iso}^{\circledast}$ in Eq.~1) rather than the original data, which have higher angular resolution. This is particularly a problem for faint compact objects imaged at a relatively shallow depth, because the object SNR will be significantly reduced due to correlated noise from smoothing.
On the other hand, the photon counting process is conducted in the areas defined by object isophotes in the NIR wavelength. The curves of growth are significantly different between filters in NIR and NUV wavelengths, with the latter much steeper, even after PSF homogenization \citep[see e.g.,][]{whitakerHubbleLegacyField2019}.
H-band isophotes therefore inevitably include a sizable fraction of pixels that are mostly noise in the NUV bandpass, as shown in Figure~\ref{fig:VtoH_illustrate}. This further decreases photometric SNRs.

As a consequence, we adopt a second option to perform UV-optimized aperture photometry, largely following the methodology of the Hubble Ultra-Deep Field UV analysis \citep[UVUDF,][]{teplitzUvudfUltravioletImaging2013,rafelskiUvudfUltravioletNearinfrared2015}.
Thanks to the CANDELS imaging campaign, most of our UVCANDELS fields have pre-existing ACS/F606W coverage, albeit with varying depth: 5600, 8600, 3300, and 5700 seconds for GOODS-N (wide region), GOODS-S (wide region), COSMOS, and EGS, respectively.
For this method, we run \textsc{SExtractor} dual-image mode using F606W (V-band) as the detection image to obtain object isophotes measured in optical wavelengths, which are smaller and much more appropriate for counting UV photons (see Fig.~\ref{fig:VtoH_illustrate}).
As provided in Table~1, the detection threshold values are slightly different across the four fields to compensate for the different exposure times to make sure identical aperture sizes are derived for objects with similar F606W surface brightness. 
In the process of detection, the V-band segmentation maps are merged and modified to follow the H-band segmentation IDs and regions (the "VtoH" segmentation; see Fig.~\ref{fig:VtoH_illustrate}). The following four scenarios are taken into account: 1) V-band isophotes without H-band counterparts are simply discarded; 2) if there are multiple V-band isophotes within one single H-band isophote, the combination of all the isophotal flux (${\rm flux}_{\rm iso}$) measurements from all those overlapping V-band segmentation regions is reported as the V-band ${\rm flux}_{\rm iso}$ for that single H-band object, with errors properly propagated; 3) H-band detected objects without V-band counterparts are assigned a null value for their V-band ${\rm flux}_{\rm iso}$; 4) if one V-band isophote contains overlapping pixels with multiple H-band segmentation regions, this single V-band ${\rm flux}_{\rm iso}$ is assigned to the H-band segmentation region with the maximum number of pixels in overlap, and a non-detection is claimed for all other H-band regions.
For the measurement images, we take advantage of the \emph{original} F275W/F435W science mosaics without PSF homogenization, following,
\begin{equation}
    {\rm mag}_{\rm tot}^{\rm improved} = {\rm mag}_{\rm iso} + (V_{\rm tot}^{\rm candels} - V_{\rm iso}).
\end{equation}
Here $V_{\rm tot}^{\rm candels}$ represents the total V-band magnitudes reported in the CANDELS photometric catalogs. ${\rm mag}_{\rm iso}$ and $V_{\rm iso}$ are the isophotal magnitudes measured from the \emph{original} high resolution data within object V-band isophotes, instead of using PSF matching and H-band isophotes which degrade image resolution and amplify noise.
In essence, we combine both the PSF and aperture corrections into one term within the parenthesis on the right hand side of Eq.~(2), given the fact that the PSF properties are similar between F275W and F606W, due to the proximity of the observed wavelengths and detector design.   
Hereafter, we refer to the measured ${\rm mag}_{\rm tot}^{\rm improved}$ as the results from the V-isophot aperture photometry.
In total, we detect 5778, 2338, 3832, and 6465 sources with SNR$\geq$3 in F275W in the GOODS-N, GOODS-S, COSMOS and EGS fields, respectively, using the V-isophot aperture photometry.
At a 5-$\sigma$ detection threshold, we obtain 3578, 1382, 2210, 3613 objects in these fields, respectively, roughly twice the yield of the H-isophot method.
Note that using V-isophot method we still report UV fluxes for all H-band selected objects from CANDELS without adding any new objects; the gain in yield at faint UV magnitudes is entirely caused by our superior photometry method.
Applying this method to the new B-band images, we measure 8515 (6062) and 14965 (10936) sources at SNR$\geq$3(5) in the COSMOS and EGS fields, respectively.

\begin{figure*}
    \centering
    \includegraphics[width=\textwidth]{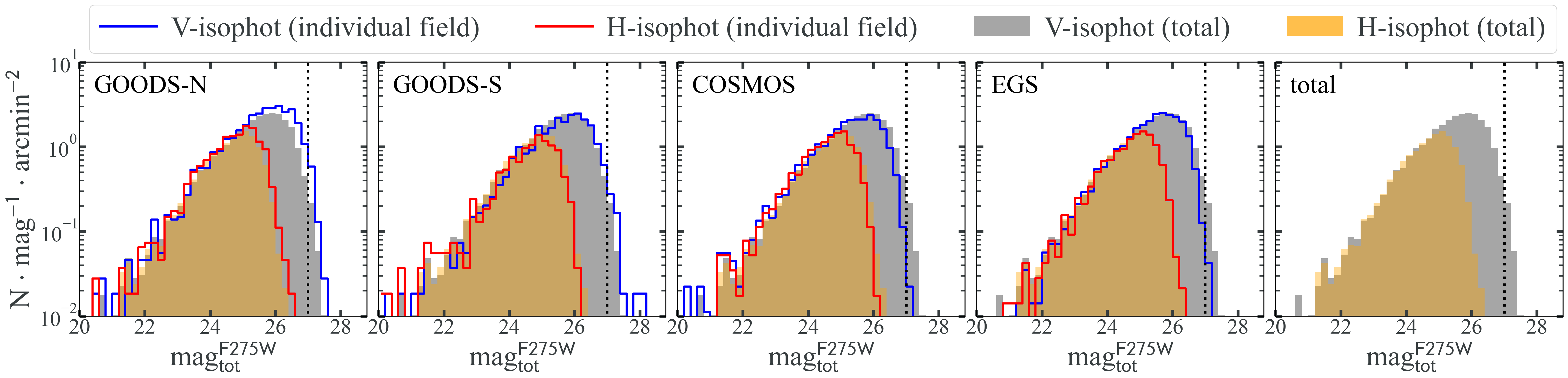}\\
    \includegraphics[width=\textwidth]{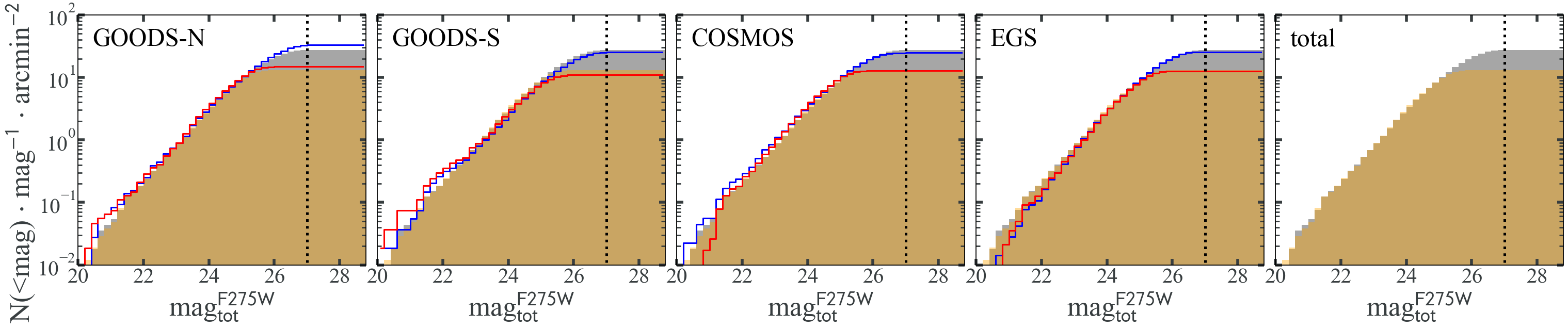}
    \caption{Differential (\textbf{top}) and cumulative (\textbf{bottom}) UV source number counts in the four individual fields and the entire data set of UVCANDELS. Here we only include sources that are detected in F275W at a significance of SNR$\geq$5. The far right panels on both rows show the number counts combined from the four fields, which are also shown in the left four panels for comparison. The results given by the H-isophot and V-isophot aperture photometry methods in individual fields are represented by the red and blue histograms, respectively. The vertical dotted line marks the expected 5-$\sigma$ limiting magnitude of 27 ABmag for point-like sources. From this, it is clearly that our V-isophot photometry is superior to the conventional CANDELS photometry method in the UV since it is capable of pushing the detection limit to fainter magnitudes. 
    }
    \label{fig:source_cnts}
\end{figure*}

The differential and cumulative source number counts in individual fields and the entire UVCANDELS dataset are displayed in Figure~\ref{fig:source_cnts}. Here we only show sources that are detected with sufficient significance (SNR$\geq$5). The 5-$\sigma$ limiting magnitude of 27 ABmag of compact sources (0.2" radius) is highlighted by the vertical dotted line, expected from the 3-orbit depth UVCANDELS F275W exposures. By performing photometry in smaller and more appropriate apertures using the original science images without degradation of image quality, our UV-optimized aperture photometry method reaches the expected depth in F275W, deeper by $\sim$1 ABmag than the depth reached by the conventional CANDELS H-isophot method.
On average, our UV-optimized photometry yields a factor of $1.5\times$ increase in SNRs in F275W, with a greater increase for brighter and more extended objects, fully realizing the potential of our modestly deep UV imaging afforded by UVCANDELS.

Henceforth, we take the F275W and F435W photometry results obtained from the V-isophot photometry as our default measurements, which complement the pre-existing CANDELS photometric catalog presented in \citet{barroCANDELSSHARDSMultiwavelength2019}, \citet{guoCANDELSMULTIWAVELENGTHCATALOGS2013}, \citet{nayyeriCANDELSMULTIWAVELENGTHCATALOGS2017}, and \citet{stefanonCANDELSMultiwavelengthCatalogs2017}.  
The detailed content of our UVCANDELS photometric catalogs is given in Table~\ref{table:phot_cat}.
These photometric catalogs will be made publicly available on the Barbara A. Mikulski Archive for Space Telescopes (MAST)\footnote{\url{https://archive.stsci.edu/hlsp/uvcandels}}, alongside the full image mosaics produced from the UVCANDELS data.

\begin{longrotatetable}
{
\begin{table*}[ht]
\caption{Description of the UVCANDELS Photometric Catalogs \label{table:phot_cat}}
\hspace*{-3cm}
\begin{threeparttable}
\begin{tabular}{llll}
\hline\hline
Column No. & Column Title & Description  & Units \\
\hline
1 & ID & object identifier in the corresponding CANDELS catalog  &  \nodata \\
2 & IAU Designation &  & \nodata \\
3--4 & RA, DEC & right ascension and declination  & decimal degrees\tablenotemark{a} \\
5 & ID\_UV & object identifier in the UVCANDELS catalog  &  \nodata \\
6 & DIST\_MATCH & cross-match distance between ID\_UV and ID in the CANDELS catalog & decimal degrees \\
7--8 & WFC3\_F275W\_FLUX\_OLD, WFC3\_F275W\_FLUXERR\_OLD & total F275W flux from the CANDELS method (H-isophot aperture photometry) & $\mu$Jy \\
9 & FLAG\_COVER & whether this object is covered (``1'') or not (``0'') by the UVCANDELS footprints & \nodata \\
10--11 & WFC3\_F275W\_FLUX\_IMPROVED, WFC3\_F275W\_FLUXERR\_IMPROVED & total F275W flux from the improved method (V-isophot aperture photometry) & $\mu$Jy \\
12 & WFC3\_F275W\_BKG\_IMPROVED & median background level within sources' combined isophotal areas for the improved method & counts per second \\
13 & WFC3\_F275W\_ISOAREA\_IMPROVED & isophotal areas above the analysis threshold for the improved method & pixel$^2$ \\
14 & WFC3\_F275W\_ISOAREAF\_IMPROVED & isophotal areas (filtered) above the detection threshold for the improved method & pixel$^2$ \\
15 & FLAG\_MULTINUM & the “multinum” flag being ``1'' or ``0'' & \nodata \\
16 & FLUX\_MAX\_OLD\_F275W & peak flux & counts per second \\
17--18 & FLUX\_ISO\_OLD\_F275W, FLUXERR\_ISO\_OLD\_F275W & isophotal flux and error & counts per second \\
19--20 & FLUX\_ISOCOR\_OLD\_F275W, FLUXERR\_ISOCOR\_OLD\_F275W & corrected isophotal flux and error & counts per second \\
21--22 & FLUX\_AUTO\_OLD\_F275W, FLUXERR\_AUTO\_OLD\_F275W & AUTO flux and error & counts per second \\
23--24 & FLUX\_PETRO\_OLD\_F275W, FLUXERR\_PETRO\_OLD\_F275W & PETRO flux and error & counts per second \\
25--26 & FLUX\_BEST\_OLD\_F275W, FLUXERR\_BEST\_OLD\_F275W & BEST flux and error & counts per second \\
27--48 & FLUX\_APER\_OLD\_F275W, FLUXERR\_APER\_OLD\_F275W & fixed aperture flux in 11 circular apertures of radius 1.47, 2.08, 2.94, 4.17, 5.88, 8.34, 11.79, \\
 &  &   16.66, 23.57, 33.34, 47.13 pixels (with plate scale being 60 mas) & counts per second \\
49 & BACKGROUND\_OLD\_F275W & background level at centroid position & counts per second \\
50 & ISOAREA\_IMAGE\_OLD\_F275W & isophotal areas above the analysis threshold for the CANDELS method & pixel$^2$ \\
51 & ISOAREAF\_IMAGE\_OLD\_F275W & isophotal areas (filtered) above the detection threshold for the CANDELS method & pixel$^2$ \\
52--54 & FLUX\_RADIUS\_OLD\_F275W & radius of 20\%, 50\%, 90\% enclosed light & pixel \\
55 & FWHM\_IMAGE\_OLD\_F275W  &  image full width half maximum & pixel \\
\hline
\end{tabular}
\begin{tablenotes}
  \item The detailed content of the UVCANDELS photometric catalogs presented in this work. The columns listed above correspond to those in
  the F275W catalogs of the four individual fields. The F435W catalogs of the EGS and COSMOS fields are also produced using the same
  method, with the names of the corresponding columns changed from F275W to F435W.
  \item a. (J2000)
\end{tablenotes}
\end{threeparttable}
\end{table*}
}
\end{longrotatetable}

\section{Sample selection}
\label{sec:select}

The wide wavelength coverage in the UVCANDELS fields (from UV to near-IR) enables high quality photometric redshift estimates (Mehta et al. in prep, following the methodology from \citealt{rafelskiUvudfUltravioletNearinfrared2015}). We thus construct galaxy samples according to their photometric redshifts. Although the full details will be presented in Mehta et al. (in prep), here we briefly describe the methods employed to derive accurate photometric redshifts.
We use the full CANDELS multi-wavelength photometric catalogs from UV to near-IR, with the new F275W and F435W data acquired by UVCANDELS, to infer photometric redshifts of UVCANDELS sources.
We calculate the photometric redshifts by combining the results from several different codes: EAZY \citep{Brammer08}, BPZ \citep{Benitez00, Coe06}, LePhare \citep{Arnouts99, Ilbert06}, and zphot \citep{Giallongo98, Fontana00}. These codes were chosen as they were consistently among the top performers in photometric redshift review papers of CANDELS fields \citep{Hildebrandt10, Dahlen13, Pacifici23}. We run two separate iterations of EAZY with multiple template sets to give us a total of 5 independent code results. 

Following the procedures outlined in \citet{Dahlen13}, we obtain the final combined results by adding the smoothed probability distributions from each code together and renormalizing it.  
Comparisons between photometric and spectroscopic redshifts for each CANDELS field present a normalized median absolute deviation \citep[NMAD,][]{Brammer08} of $\sigma_\text{NMAD}$ = $0.0193, 0.0303, 0.0356$ and $0.0260$ in COSMOS, EGS, GOODS-N and GOODS-S, respectively, and a outlier fraction of $\eta$ = $1.32\%, 1.63\%, 1.08\%$ and $1.15\%$, respectively (see also Table~\ref{tab:4fields}). For the entire sample combining all four UVCANDELS fields, our photometric redshift analysis achieves $\sigma_\text{NMAD}=0.0263$ and $\eta=1.32\%$.

Aiming to measure the UV LF, we choose the absolute magnitude at commonly used rest-frame 1500\AA\ as a UV magnitude indicator and use F275W to measure it in this work. The corresponding redshift range can then be defined to be $0.6<z<1$ according to the F275W wavelength range of [2286\AA, 3120\AA]. 
We consider only those sources that have the V-isophot aperture photometry measurements in the following LF analysis. This leads to a slightly smaller survey area of $\sim 392$ arcmin$^2$ in total, due to the lack of V-band coverage in parts of the COSMOS field. The specific survey area of each field used in our analysis are tabulated in Talbe~\ref{tab:4fields}. 
Besides the advantage of the $\sim1$ magnitude improvement of the survey depth (at $5\sigma$ level) as discussed in Section~\ref{sec:photom}, using only the V-isophot aperture photometry results can also help in conducting a self-consistent completeness analysis (see Section~\ref{sec:comp}). 
In order to avoid selecting spurious objects, we require 3$\sigma$ measurements in the rest-frame UV filter. The selection criteria for our samples are:
\begin{enumerate}[itemsep=0mm,label=\alph*.]
    \item $0.6<z_{\text{phot}}<1$   
    \item ${\rm SNR}>3$ in the F275W band.
\end{enumerate}
This results in a selection of 1361, 2001, 1823 and 696 galaxies in the COSMOS, EGS, GOODS-N and GOODS-S fields, respectively. 

Furthermore, we examine and exclude possible sources of contamination to ensure the purity of our samples. Firstly, we perform cross-match analysis on our sample with the CANDELS AGN catalogs compiled by Kocevski et al. based on x-ray observations (private communication), removing 4, 25, 29 and 12 matched sources in COSMOS, EGS, GOODS-N and GOODS-S fields, respectively. The number of X-ray sources rejected varies notably across the 4 fields, due to different survey area of each field and depths of the x-ray observations. Secondly, we implement a star/galaxy separation criteria on the UVCANDELS F606W (the detection band in our V-isophot photometry) catalog, as follows:
\begin{enumerate}[itemsep=0mm,label=\alph*.]
\item ${\rm SNR}>3$      
\item $\text{CLASS}\_\text{STAR}>0.9$      
\item $\text{elongation}=\text{A\_IMAGE}/\text{B\_IMAGE}<1.2$      
\item $f_{\rm 3~arcsec}/f_{\rm 0.7~arcsec}<1.3$    
\item $\text{m}_{\text{F606W}}<22\ \text{ABmag}$.
\end{enumerate}
The sources that meet this criteria are identified as stars and rejected from our sample.

Finally, at the bright end of $M_{\rm UV}< -20$, we perform a visual inspection to further exclude some spurious objects, such as wrongly segmented sources. We find that with the improved (V-isophot) photometry method, those falsely detected bright sources with the old (H-isophot) photometry method (e.g., due to segmentation fault) are now almost all automatically eliminated, which further validate our UV-optimized aperture photometry method. After removing all the contaminants mentioned above, there were 1357, 1976, 1794 and 683 galaxy candidates selected from the four fields respectively, summing up to 5810 sources appropriate for our UV LF analysis. We listed these numbers also in Table~\ref{tab:4fields}.

The rest-frame absolute magnitude at 1500\AA\ of the selected galaxy are then determined by \citep[e.g.,][]{Hogg02}
\begin{equation}
M_{1500}=m_{\mathrm{F275W}}-5\text{log}(d_{L}/10\text{pc})-K_\text{cor}
\label{eq:M1500}
\end{equation}
\noindent where $K_\text{cor}$ represents the K-correction from the emission waveband centered around rest-frame 1500\AA\ to the observed band F275W. By limiting our measurements to a redshift range ($0.6<z<1$) in which the rest-frame emission is red-shifted to and observed through the selected filter, the k-correction term is minimized. We estimate $K_\text{cor}$ for each individual galaxy in our catalog via interpolation on a pre-computed $K_\text{cor}(z, \beta)$ table, where $\beta$ is the UV continuum slope, rather than performing the full integration of the best-fit spectrum. Here the $K_\text{cor}(z, \beta)$ table is computed using equation(13) in \citet{Hogg02} given a model spectrum consisting of a power law $F(\lambda)\sim\lambda^\beta$ and a Lyman break at 1216\AA. The specific $\beta$ value for each galaxy is obtained based on their spectral energy distribution (SED) fitting results (Mehta et al. in prep). We ensure that uncertainties on $M_{1500}$ originated from our k-correction calculations are below the level of uncertainties due to flux and redshift measurement errors.

\begin{figure}
\centering
\includegraphics[trim=4cm 11cm 4cm 10cm, width=0.45\textwidth]{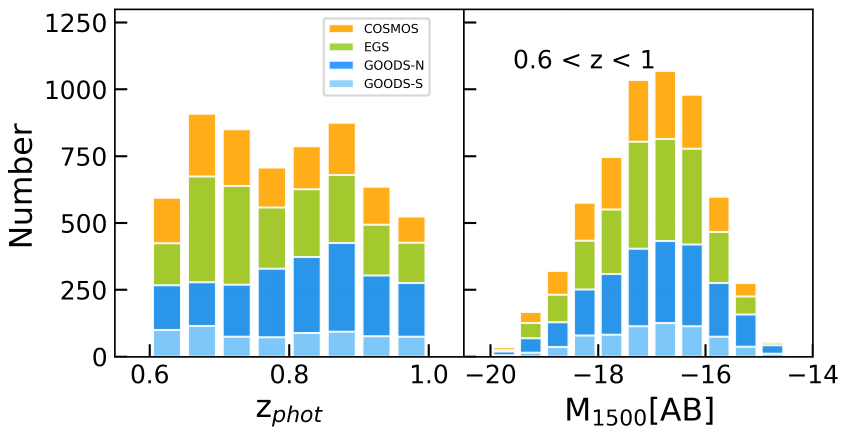}
\caption{The photometric redshift (left panel) and rest-frame 1500 \AA\ absolute magnitude (right panel, measured from F275W) distributions of the selected 5810 sources from the UVCANDELS photometry catalogs at $0.6<z_{\rm phot}<1$ that have ${\rm SNR}\geq3$ for F275W flux. Possible contamination from stars, AGNs and suprious objects have been examined and excluded.} 
\label{fig:bin1-z-hist}
\end{figure}

Figure~\ref{fig:bin1-z-hist} shows the redshift distribution (left panel) of these 5810 selected galaxies, which is a relatively flat distribution across the redshift range of $0.6<z<1$. The median redshift is $\sim0.8$. The distribution of rest-frame $M_{1500}$ of our sample is also given in Figure~\ref{fig:bin1-z-hist} (right panel), with a faint end magnitude limit down to $-14.2$.

\begin{deluxetable}{lllll}
\tablecaption{Description of selected F275W samples from 4 UVCANDELS Fields\tablenotemark{a}}
\tablehead{ 
\colhead{Field} &
\colhead{Area\tablenotemark{b}} & \colhead{N\tablenotemark{c}} & \colhead{$\sigma_\text{NMAD}$} & \colhead{$\eta$} 
}
\startdata
COSMOS  & 89.12  & 1357 & $0.0193$  & $1.32\%$ \\
EGS     & 141.14 & 1976 & $0.0303$  & $1.63\%$ \\
GOODS-N & 107.92 & 1794 & $0.0356$  & $1.08\%$  \\
GOODS-S & 54.13  & 683  & $0.026$   & $1.15\%$  \\
\hline
All     & 392.31 & 5810 & $0.0263$  & $1.32\%$
\enddata
\tablenotetext{a}{F275W exposures at a
uniform 3-orbit depth, except that in GOODS-N, CVZ increased the efficiency of the observations.}
\tablenotetext{b}{Covered both by F275W(for UV LF measurements) and F606W(for V-isophot aperture photometry), in units of arcmin$^2$.}
\tablenotetext{c}{Number of galaxies selected with ${\rm SNR}\geq3$ in F275W and photometric redshift $0.6<z<1$, after removing contamination. }
\label{tab:4fields}
\end{deluxetable}

\section{Completeness analysis}
\label{sec:comp}

Galaxy imaging surveys suffer from incompleteness where not all sources of a targeted population can be universally observed, due to defects in observations, instruments, data reduction and sample selection. To encompass the volume density properly in computing UV LF, one therefore needs to estimate the incompleteness of a sample and correct it precisely, which is more critical for the faint end approaching the survey detection limit. A commonly used procedure to estimate the completeness in galaxy surveys is to inject artificial galaxies with properties similar to the underlying population into real images, then conduct identical data reduction and sample selection as the observed sample, and finally calculate the fraction of recovered mock galaxies as a function of magnitude, redshift, galaxy size, etc \citep[e.g., ][]{Oesch10,Alavi16}.

As mentioned in Section~\ref{sec:photom}, UVUDF shares a similar instrument, targeted galaxy population, data reduction and photometry pipeline with UVCANDELS. Therefore, the completeness simulation results for UVUDF \citep{mehtaUVUDFUVLuminosity2017}, which were computed following the standard mock-source injection and recovery technique, should be applicable to UVCANDELS. A primary difference between the two surveys related to sample completeness is that the limiting magnitude of UVUDF ($m_{\rm F275W}=27.8$) is 0.8 magnitude deeper than that of UVCANDELS ($m_{\rm F275W}=27$). Given that the sample completeness is a function of SNR, we can adapt the completeness function derived for UVUDF to that suitable for UVCANDELS data by adjusting the SNR threshold to account for the different survey depths.

\begin{figure}
\centering
\includegraphics[width=0.45\textwidth]{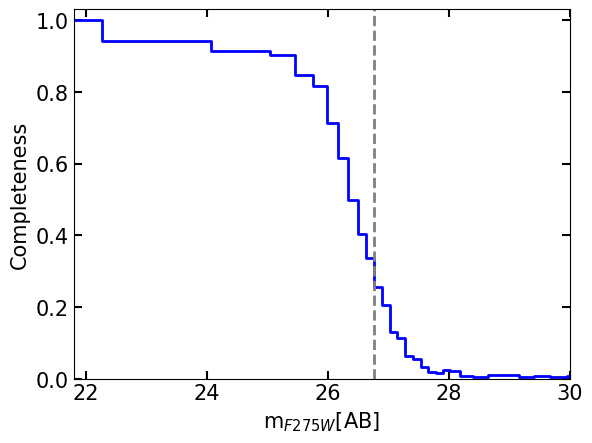}
\caption{Completeness as a function of apparent magnitude in F275W, $C(m_{\mathrm{F275W}})$, for our UVCANDELS sample, as adapted from UVUDF \citep{mehtaUVUDFUVLuminosity2017}. The vertical dashed line marks the limiting magnitude determined by the $30\%$ completeness.}
\label{fig:c-m}
\end{figure}

Figure~\ref{fig:c-m} shows the completeness as a function of apparent magnitude in F275W, $C(m_{\mathrm{F275W}})$, for our sample, adapted from UVUDF simulation results. To ensure reliable incompleteness corrections, particularly at the faint end, following \citet{mehtaUVUDFUVLuminosity2017} we limit the galaxy sample for our LF analysis where the completeness is higher than $30\%$. We do not apply a higher limit of completeness cut here since we intend to study the faint end of the LF. This provide a final sample of 4726 galaxies used to derive the UV LF. The vertical dashed line in Figure~\ref{fig:c-m} shows the relevant faint end magnitude limit, $m=26.77$. We also compute the completeness as a function of redshift and UV absolute magnitude $C(z,M)$ by sampling on a z-grid, which will be used in the following subsection to define the effective survey volume.

Despite the employment of the V-isophot photometry, our UV sample is built intrinsically based on the H-band detection due to the fact that registration of sources is done in H-band. This may give rise to a color-dependent incompleteness since sources that are blue enough to be detected in F275W whilst fainter than the limiting magnitude of F160W ($\sim27$AB) will be omitted. Resorting to the V-band detection catalog, we evaluate the fraction of sources suffering from this effect. As a result, $\sim3\%$ UV sources are found bright with $m_\text{F275W}<27$AB at a $S/N>3$ level meanwhile dropping out in H-band detection.  
Moreover, we also estimate the fraction of galaxies with a best-fit color $\mathrm{F275W}-\mathrm{F160W}<0$ using the SED catalog. Only $0.28\%$ sources are found of this color among those above the detection limit of F160W. Overall, uncertainty from this incompleteness effect turns out to be smaller than the Poisson noise on galaxy number density per magnitude bin, and hence its impact on our primary results should be negligible.

\subsection{The Effective Survey Volume}
\label{subsec:volume}

The incompleteness of a sample affects the number counts of sources given the survey volume. Or equivalently, it impacts the effective survey volume of a sample, which is critical in the computation of the LFs. Hence we incorporate the completeness corrections in the computation of the effective survey volume ($V_{\mathrm{eff}}$) as follows,
\begin{equation}
V_{\mathrm{eff}}(M)=\int_{z_{\mathrm{min}}}^{z_{\mathrm{max}}} \frac{dV_{\mathrm{com}}}{dz\ d\Omega}\ C(z,M)\ \Omega(z) \ \mathrm{d}z
\label{eq:v_eff}
\end{equation}
where $z_{\mathrm{min}}$ and $z_{\mathrm{max}}$ are the lower and upper redshifts of our selected sample, respectively. $dV_{\mathrm{com}}$ is the comoving volume element per unit area $d\Omega$ at a redshift $z$. $C(z,M)$ is the completeness function depending on redshift $z$ and UV absolute magnitude $M$. $\Omega(z)$ is the survey area at $z$. For our sample with the improved photometry measurements, the corresponding survey area is $\sim 392$ arcmin$^2$ in total, a constant throughout the redshift range $0.6<z<1$.

\section{Luminosity function}
\label{sec:lf}

The galaxy LF is well fit by a parameterized Schechter function \citep{Schechter76}, a power-law with slope $\alpha$ at the faint-end and an exponential cut-off at luminosities brighter than a characteristic magnitude, $M^{*}$, as below\footnote{In statistics this is recognised as the probability density function of the gamma distribution.}:
\begin{equation}
\phi(M)=0.4\ \mathrm{ln}(10)\ \phi^{*}\ 10^{-0.4(M-M^{*})(1+\alpha)}\ e^{-10^{-0.4(M-M^{*})}}
\label{eq:sch}
\end{equation}
where $\phi^{*}$ is the normalization factor.

Using the effective survey volume accounting for the completeness corrections, we can derive the rest-frame 1500 \AA\ UV LFs of our sample. We first compute the binned UV LFs, then we perform a maximum likelihood estimate on the unbinned data to find the best-fit Schechter parameters.

\subsection{The Binned UV LFs}
\label{subsec:binned lf}

Given the measured effective survey volume $V_\mathrm{eff}$, the LF value at each absolute magnitude bin can be calculated using the commonly used $V_\mathrm{eff}$ method \citep[e.g.,][]{Steidel99, Alavi14}. In this method, the number density of galaxies in each absolute magnitude bin is quantified by dividing the number of galaxies by the effective volume covered by that bin. The effective volume, however, might vary significantly from one side of the magnitude bin to the other. Following \citet{Alavi16}, we therefore calculate the effective volume for each individual galaxy and then sum up over all galaxies within each bin, as below:
\begin{equation}
\phi(M_{i})dM_{i}=\sum_{j=1}^{N_i}\frac{1}{V_{\mathrm{eff}}(M_{j})}
\end{equation}
where $N_i$ is the total number of galaxies in the $i$th bin, and $V_\mathrm{eff}$ is the effective volume covered by a galaxy with absolute magnitude $M_j$ in the $i$th bin, summing up over all 4 fields. 

A Poisson error, $\frac{\phi_{i}}{\sqrt{N_i}}$, is assigned for each bin where the number of galaxies $N_i$ is larger than 50. For those bins with $N_i<50$, we use the Poisson approximation, $\Delta_{P}$, from \citet{Gehrels86} to determine the uncertainty as $\frac{\phi_{i} \Delta_{P} }{N_i}$. The bin width is set to be $\Delta M_{UV}=0.5$ magnitude. The faintest magnitude bin is centered at $M_{UV}=-15.75$, after the completeness cut of $>30\%$ is implemented. Since both the redshift ranges and the luminosity bins considered in this analysis are relatively small, $V_\mathrm{eff}$ of different sources within each luminosity bin do not vary significantly and consequently the weights of sources are similar. Therefore the Poisson error should still be a good estimate here.

We emphasize that the binned LF provided in Figure~\ref{fig:lf-bin1} is simply for visual inspection and comparison. Because of arbitrary bin widths, bin centers and loss of information within each bin, it is not a good choice to use the binned data for parameter inference. We therefore perform a maximum likelihood estimate on the unbinned data discussed in the next section. 

\subsection{The Unbinned Maximum Likelihood Estimator}
\label{subsec:mle}

In this section, we perform a maximum likelihood estimate (MLE) to infer the best fit Schechter parameters of the UV LFs, using the unbinned data. Specifically, we adopt the modified form of MLE \citep{Alavi16} to account for the measurement errors of the absolute magnitude, with similar forms also presented in \citet{Alavi14}, \citet{Mehta15} and \citet{mehtaUVUDFUVLuminosity2017}. In this scenario, the best fit parameter values are found by maximizing the joint likelihood function of all galaxies as shown below:
\begin{equation}
    \mathcal{L}=\prod_{i=1}^{N}P(M_{i})
    \label{eq:L}
\end{equation}
where N is the total number of objects down to the magnitude limit of the sample. $P(M_{i})$ is the probability of detecting a galaxy with absolute magnitude $M_{i}$, defined as: 
\begin{equation}
    P(M_{i})=\frac{\int_{-\infty}^{+\infty} \phi(M) \ V_{\mathrm{eff}}(M)\ G(M|M_{i},\sigma_{i}) \ \mathrm{d} M}{\int_{-\infty}^{M_{\mathrm{limit}}} \phi(M) \ V_{\mathrm{eff}}(M)\ \mathrm{d} M}
\label{eq:pm_mod}
\end{equation}

\noindent Here, $\phi(M)$ is the LF given by Equation~\ref{eq:sch} and $V_{\mathrm{eff}}(M,z)$ is the effective volume. $M_\mathrm{limit}$ is defined to be the faintest absolute magnitude of a sample, with $M_\mathrm{limit}=-15.52$ for our sample. $G(M|M_{i},\sigma_{i})$ is a Gaussian probability distribution assumed for each object to include the uncertainties of its absolute magnitude: 

\begin{equation}
    G(M|M_{i},\sigma_{i})=\frac{1}{\sqrt{2\pi}\sigma_{i}}\text{exp}(-\frac{(M-M_{i})^{2}}{2\sigma_{i}^{2}})
\end{equation}
where the total uncertainty, $\sigma_{i}$, of the absolute magnitude is obtained by adding the photometry uncertainty $\sigma_{m}$ and the photometric redshift uncertainty $\sigma_{z}$ in quadrature. The photometry uncertainty is computed using the \texttt{SExtractor} output of flux error. The photometric redshift uncertainty on the measured absolute magnitude of each galaxy is calculated using the $1\sigma$ confidence interval of its redshift probability distribution and error propagation of Equation~\ref{eq:M1500}. 

We note that, with the MLE fitting technique, the normalization parameter of the Schechter function is cancelled out and not fitted. Therefore, we must estimate it separately from the number counts (e.g., \citet{Alavi14}),
\begin{equation}
    \phi^{*}=\frac{N}{\int_{M_{1}}^{M_{2}} \phi(M) \ V_{\mathrm{eff}}(M)\ \mathrm{d}M}
\end{equation}
Where $M_{1}$ and $M_{2}$ represent the brightest and faintest objects in the sample, respectively.

We compute the probability function for each individual galaxy in our sample and estimate the best fit Schechter parameters by maximizing the joint likelihood function of all galaxies. Flat priors of $-3<\alpha<0$ and $-30<M^{*}<-14$ are imposed throughout.
We perform a Markov Chain Monte Carlo analysis (MCMC) to quantify the uncertainties of our best-fit parameters using the Python package \texttt{emcee} \citep{Foreman13}.

\subsection{Results}
\label{subsec:results}

Along with the binned UV LF, we present the best-fit UV LF using the MLE techniques described above for our photometric redshift selected sample. As discussed in Section~\ref{sec:comp}, we impose a completeness cut of $>30\%$ on our sample to avoid using objects with completeness corrections that are too large. This leads to a final sample size of 4726 galaxies used to derive the UV LF and a faint end magnitude limit $M_\mathrm{limit}$ of $-15.52$, which is also listed in Table~\ref{tab:pars}.  

\begin{figure*}
\centering
\includegraphics[trim=0cm 1.5cm 0cm 2.9cm,clip=true,width=\textwidth]{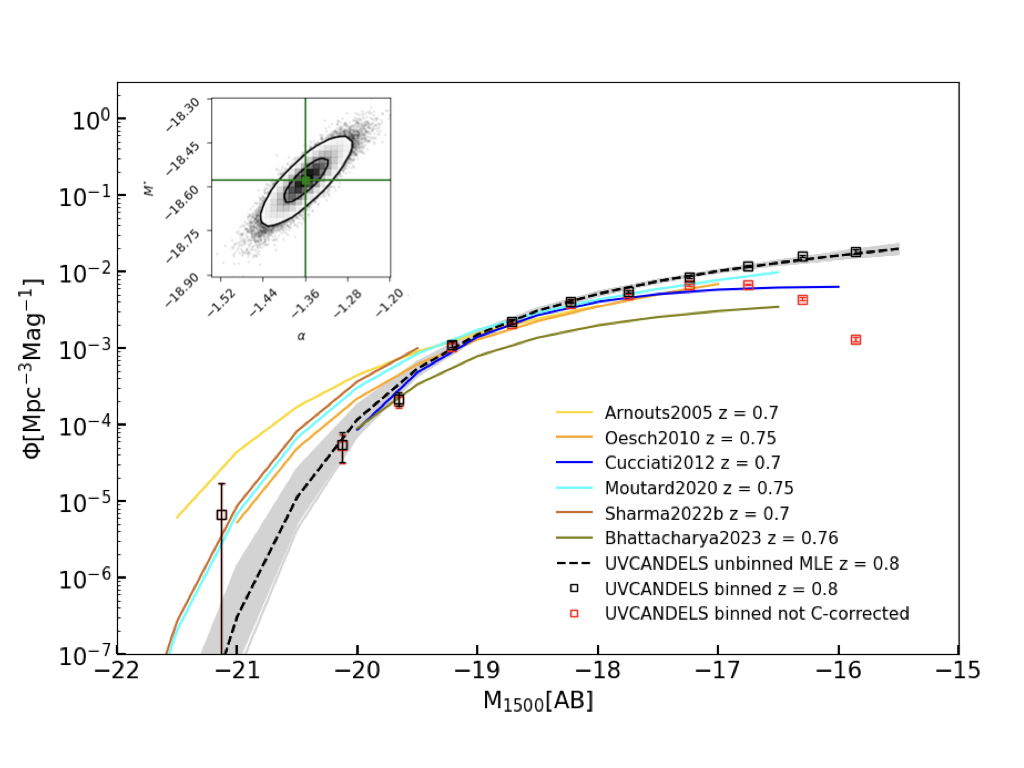}
\caption{Rest-frame UV LF of UVCANDELS galaxy at $z\sim0.8$. The black dashed line is the best-fit Schechter LF from the maximum likelihood estimate, with its $3\sigma$ errors denoted by the gray shaded region. The black/red squares represents the binned LF with/without the completeness corrections. The colored lines are results from the literature at similar redshifts summarized in the legend. The inset shows the the maximum likelihood estimates of the Schechter LF parameters $\alpha-M^{*}$. The green lines label the best-fit (median) values of $\alpha$ and $M^{*}$, reported in Table~\ref{tab:pars}. The black contours show the $1\sigma$(0.393) and $2\sigma$ ($0.865$) confidence levels.}
\label{fig:lf-bin1}
\end{figure*}

Figure~\ref{fig:lf-bin1} shows the rest-frame UV LF of the full UVCANDELS galaxy sample at a median redshift $z\sim0.8$. The black dashed line is the best-fit Schechter LF from the maximum likelihood estimate, with its $3\sigma$ errors denoted by the gray shaded region. Here we use the bootstrapped MCMC sample of $(\alpha,M^{*})$ pair to estimate the distribution of LF at each $M_\mathrm{UV}$ and calculate the corresponding uncertainty.  
The binned LF with the completeness corrections is shown by the black squares, whereas its counterpart that does not incorporate the completeness corrections is presented by the red squares. The error bars of the 2 binned LFs represent their Poisson errors as discussed in Section~\ref{subsec:binned lf}. It is noted that our best-fit LF to the unbinned data using MLE is in good agreement with the binned LF. In Figure~\ref{fig:lf-bin1}, we also plot the best-fit rest-UV LFs from the literature that are derived at the same wavelength ($\sim$1500\AA) and similar redshifts using data from GLAEX \citep{Arnouts05}, HST ERS \citep{Oesch10}, VVDS \citep{Cucciati12}, CLAUDS and HSC-SSP \citep{Moutard20} and XMM-COSMOS \citep{Sharma22b}. It is noted that our LF measurements cover a wider absolute magnitude range ($-21.5<M_\text{UV}<-15.5$) at relevant redshifts than the previous studies shown in Figure~\ref{fig:lf-bin1}. 

The specific best-fit values of Schechter parameters from our full sample are tabulated in Table~\ref{tab:pars}. Uncertainties and correlations of the faint-end slope $\alpha$ and the characteristic luminosity $M^{*}$ are illustrated by the inset contours of Figure~\ref{fig:lf-bin1}. Our MLE estimates find the faint-end slope $\alpha = -1.359^{+0.041}_{-0.041}$ at $z\sim0.8$. This value is in good agreement with \cite{Sharma22b} who reported a measurement $\alpha = -1.37^{+0.048}_{-0.043}$ at $z\sim0.7$, and is relatively flat than other studies at similar redshifts especially those estimated under brighter magnitude limit, such as \citet{Arnouts05} and \citet{Oesch10}, whereas we are still in consistency with them within $3\sigma$ considering their large error bars. We note that most of the studies listed in Figure~\ref{fig:lf-bin1} assume a flat LCDM with $\Omega_m=0.3$ and $h=0.7$. If the same cosmology is adopted, we obtain $\alpha = -1.409^{+0.042}_{-0.041}$, with a deviation $\delta\alpha=-0.05$ comparable to the $1\sigma$ statistical uncertainty. Deviation of $M^{*}$ is even much smaller, with $\delta M^{*}=-0.003$ compared to the constraint $M^{*}=-18.578^{+0.073}_{-0.075}$ based on the adopted cosmology throughout this paper. Therefore the effect of assumption of different cosmological models on our LF measurements is negligible.

The Poisson errors on the number counts are already accounted for by the error bars shown in Figure~\ref{fig:lf-bin1}. On the other hand, the number counts are also prone to errors due to fluctuations of large-scale structure, namely, cosmic variance. We estimate the cosmic variance for our sample using the Cosmic Variance Calculator v1.03\footnote{\url{https://www.ph.unimelb.edu.au/~mtrenti/cvc/CosmicVariance.html}} \citep{Trenti08}. We assume $\sigma_8=0.8$ and an average halo occupation fraction of $0.5$, and use the \citet{Sheth99} bias formalism. For a field-of-view of the 4 fields combined, we estimate a fractional error of $0.18$ on the number counts of bright ($M_{UV} < -20$) sources in our sample, which is smaller than the relative Poisson uncertainty of $0.34$. 

At the bright end, it is seen that there exists relatively larger difference between our LF and those from other studies as shown in Figure~\ref{fig:lf-bin1}. First, the cosmic variance mentioned above can partly explain this difference, as \citet{Arnouts05}, \citet{Moutard20} and \citet{Sharma22b} possess larger survey areas about $1.5-20$ deg$^2$, whereas this analysis has $\sim400$ arcmin$^2$. Second, we carefully remove point-like sources including AGNs, whereas \citet{Moutard20} noticed that the bright end of their LFs may suffer from contamination of stars and QSOs and in \citet{Oesch10} the AGN removal was not mentioned specifically. On the other hand, the higher resolution of HST WFC3 compared to that of GALEX \citep{Arnouts05}, CLAUDS \citep{Moutard20} and XMM-OM \citep{Sharma22b} may help in better discriminating contamination including AGNs. As a test, we obtain a good agreement ($<3\sigma$) with \citet{Oesch10}, \citet{Moutard20} and \citet{Sharma22b} at $M_{UV}<-20$ by retaining the sources that are identified as AGNs in our catalog. Besides, we also perform a visual inspection at $M_\mathrm{UV}<-20$ and carefully exclude contamination from spurious objects, including those wrongly segmented galaxies since the estimate of photometric redshift for these galaxy fragments should not be accurate. This treatment further reduces the number count of our sample at the bright end, which tends to infer a larger $M^{*}$ and might also help explain the discrepancy of the LFs at the bright end between our and other studies. 

\begin{deluxetable*}{lllllll}
\tabletypesize{\footnotesize}
\tablecaption{Best-fit Schechter Parameters for UV LFs and the UV Luminosity Density}  
\tablehead{ 
\colhead{Redshift} & \colhead{$M_{lim,\mathrm{UV}}(\mathrm{Mag})$} & \colhead{$N$\tablenotemark{a}} & \colhead{$\alpha$} & \colhead{$M^{*}(\mathrm{Mag})$} & \colhead{$\phi^{*} (10^{-3}\mathrm{Mpc}^{-3})$} & \colhead{$\rho_{\mathrm{UV}}$\tablenotemark{b}}
}
\startdata
\multicolumn{7}{c}{Full Sample}\\
\hline
$0.6<z<1.0$ & -15.52 & 4726 & $-1.359^{+0.041}_{-0.041}$ & $-18.578^{+0.073}_{-0.075}$ & $8.21^{+0.82}_{-0.78}$  & $1.339^{+0.027}_{-0.030}$ \\
\hline
\multicolumn{7}{c}{Sub-samples}\\
\hline
$0.6<z<0.8$ & -15.52 & 2482 & $-1.322^{+0.057}_{-0.056}$ & $-18.492^{+0.101}_{-0.107}$ & $8.95^{+1.24}_{-1.17}$ & $1.286^{+0.032}_{0.035}$ \\
$0.8<z<1.0$ & -16.17 & 2346 & $-1.419^{+0.067}_{-0.066}$ & $-18.676^{+0.108}_{-0.116}$ & $7.49^{+1.21}_{-1.13}$  & $1.461^{+0.059}_{-0.070}$
\enddata
\label{tab:pars}
\tablenotetext{a}{Sample size after removing sources with completeness $< 30\%$. }
\tablenotetext{b}{in units of $\times 10^{26}$ ergs/s/Hz/Mpc$^3$}
\end{deluxetable*}

\section{Discussion}
\label{sec:discussion}

\subsection{Evolution of the Schechter Parameters}
\label{sec:split-lf}

\begin{figure}
\centering
\includegraphics[width=\columnwidth]{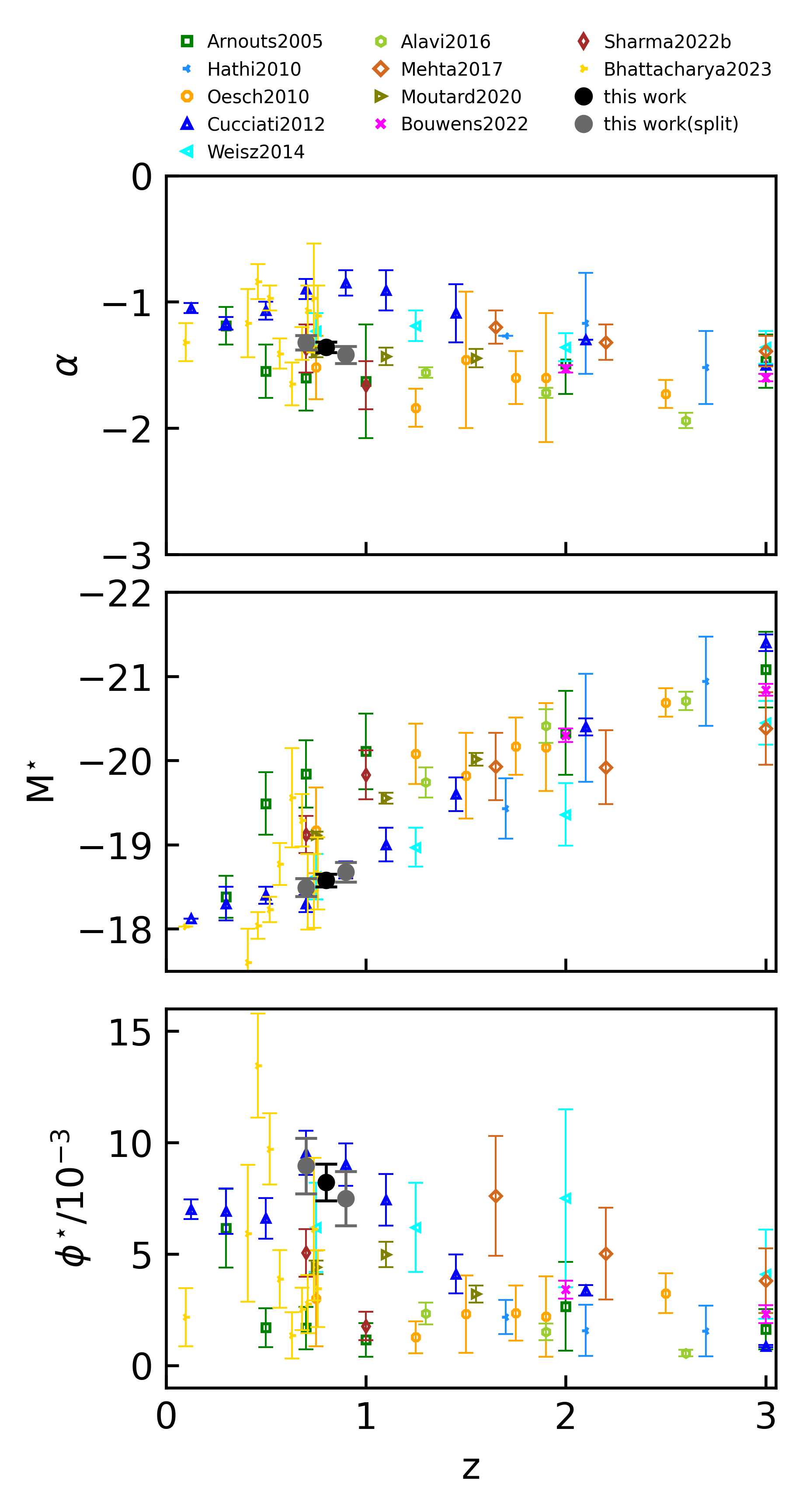}
\caption{Redshift evolution of the Schechter LF parameters $\alpha$(top), $M^{*}$(middle), and $\phi^{*}$ (bottom). The black dots shows the main results from our full sample at $z\sim0.8$. The gray dots denote the determinations of our split sub-samples at $z\sim0.7$ and $z\sim0.9$, respectively. Symbols representing previous determinations from the literature are summarized in the legend. The error bars show the $1\sigma$ uncertainties.}
\label{fig:par-evol}
\end{figure}

To investigate the evolution of the LF parameters with redshift, we illustrate determinations of the rest-frame UV LFs across the redshift range $0<z<3$ from the literature \citep{Arnouts05,Oesch10,Cucciati12,Weisz14,Alavi16,mehtaUVUDFUVLuminosity2017,Moutard20,Bouwens22,Sharma22b}, along with our best-fit parameters in Figure~\ref{fig:par-evol}. Here all results are derived at $\sim1500$\AA, except that studies of \citet{Bouwens22} are given at $1700$\AA.
The statistical errors of our best-fit parameters in Figure~\ref{fig:par-evol} are significantly smaller than those of other studies at similar redshifts, as expected from the unprecedentedly large sky coverage at high angular resolution afforded by UVCANDELS. Our statistical precision reaches a high level, with a $1\sigma$ error of $\sigma \alpha=0.041$ and $\sigma M^{*}=0.075\mathrm{Mag}$, although there may also exist systematic deviations due to, e.g., uncertainties of completeness corrections especially on low luminosity sources.

By comparing our results with those at lower and higher redshifts in Figure~\ref{fig:par-evol}, our best-fit faint-end slope $\alpha$ (in the upper panel) and characteristic $M^{*}$ (in the middle panel) are found to be consistent with other determinations at similar redshifts. We do not put emphasis on the comparison of the characteristic number density $\phi^{*}$ with previous results from the literature, since with our MLE fitting technique $\phi^{*}$ is cancelled out and not fitted directly. Overall, our estimated Schechter parameters at $z=0.8$ are in better agreement with the results at $z=0.75$ from \citet{Weisz14}, although their best-fit is derived from local group fossil records at the very faint-end of $M_{UV}>-14$. We are also in good agreement with \citet{Bhattacharya23} given their large uncertainties, especially we both infer a relatively larger $M^{*}\sim-18.5$ at $z\sim0.75-0.8$. We should note that their results are also based on UV observations in one of the four UVCANDELS fields, i.e., GOODS-N, using Astrosat\citep{Singh14} and HST as well. Combined with constraints from different redshifts, $\alpha$ exihibts an evolutionary trend of getting steeper with redshift. 

\begin{figure*}
\centering
\includegraphics[width=\columnwidth]{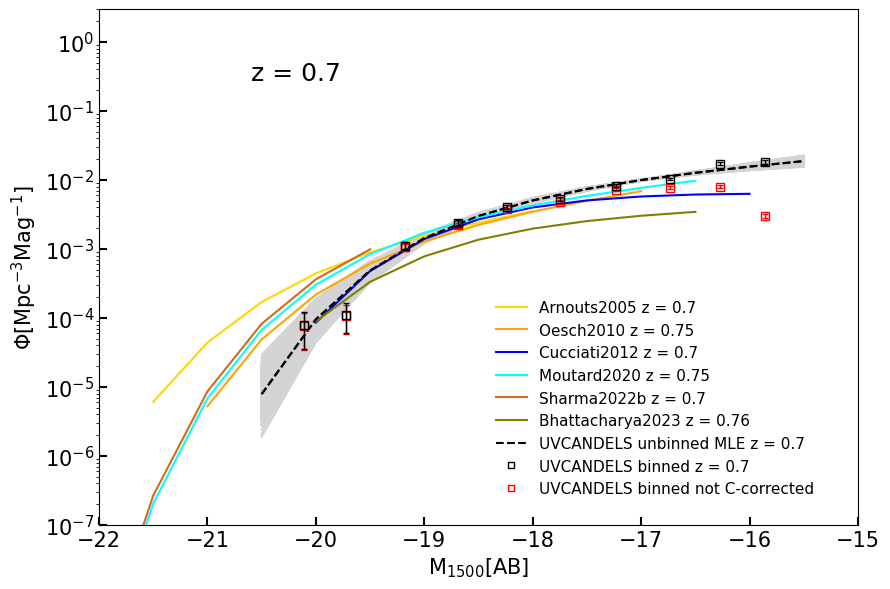}
\includegraphics[width=\columnwidth]{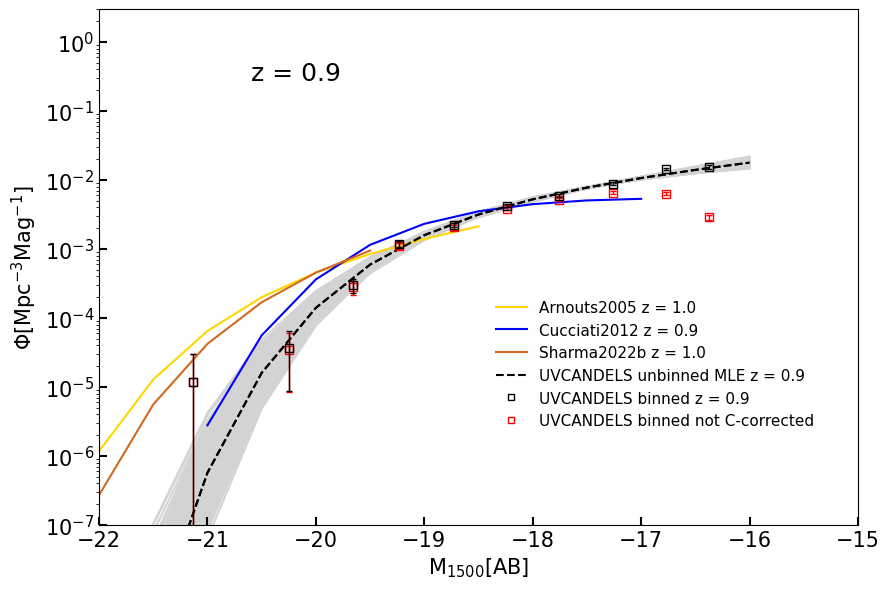}
\caption{Rest-frame UV LFs of the split sub-samples at $z\sim0.7$ (Left panel) and $z\sim0.9$ (Right panel). The black dashed line is the best fit Schechter LF using the MLE, with its $3\sigma$ errors denoted by the gray shaded region. The black/red squares represents the binned LFs with/without the completeness corrections. The colored lines are results from the literature summarized in the legend.}
\label{fig:lf-spliting}
\end{figure*}

Thanks to the large sample size and precise photometric redshifts of our sample, it is statistically feasible to split the full sample into 2 sub-samples in redshift ranges $0.6<z<0.8$ and $0.8<z<1.0$, respectively, in order to further investigate the evolution of the LF. The resulting numbers of candidates are $2482$ at mean redshift of $z\sim0.7$ and $2346$ at $z\sim0.9$. Figure~\ref{fig:lf-spliting} illustrates the best-fit UV LFs using the maximum likelihood estimates, with the relevant Schechter parameters listed in Table~\ref{tab:pars}.
With a roughly $50\%$ degradation of the uncertainties compared to the case with the full sample, the best-fit values of faint-end slope $\alpha$ are $-1.322^{+0.057}_{-0.056}$ at $z\sim0.7$ and $-1.419^{+0.067}_{-0.066}$ at $z\sim0.9$, which is also illustrated by the gray dots in the upper panel of Figure~\ref{fig:par-evol}. Although the $\alpha$ values at the two redshifts are not significantly distinguishable at a $\geq3\sigma$ level, we observe a possible evolution that the faint-end slope of the UV LF is getting steeper with redshift, which is generally consistent with studies from the literature. Combining our determinations with previous studies on a larger baseline of redshift, this evolving trend of $\alpha$ can be more clearly observed, which provides further support to the link between the build-up of galaxies and their dark matter halos (See also \citealt{Bouwens21, Bouwens22}). 
We also find that, from the middle panel of Figure~\ref{fig:par-evol}, the characteristic $M^{*}$ is slightly brighter at redshift $z\sim0.9$ relative to $z\sim0.7$ by $\Delta M^{*}\sim-0.18$, although the values are consistent with each other within $2\sigma$. 

\subsection{UV Luminosity Density}
\label{sec:uvld}

The faint-end slope of the UV LFs determines the relative contribution of faint and bright galaxies to the total cosmic UV luminosity. We use our best-fit Schechter LF in Section~\ref{sec:lf} to compute the unobscured (i.e., not corrected for dust) cosmic UV luminosity density $\rho_\text{UV}$ as:

\begin{equation}
\label{eq:uvld}
\rho_\text{UV} = \int_{L_\text{lim}}^{\infty} L\ \phi(L)\ dL = \int_{-\infty}^{M_\text{lim}} L(M)\ \phi(M)\ dM
\end{equation}

\begin{figure}
\centering
\includegraphics[width=\columnwidth]{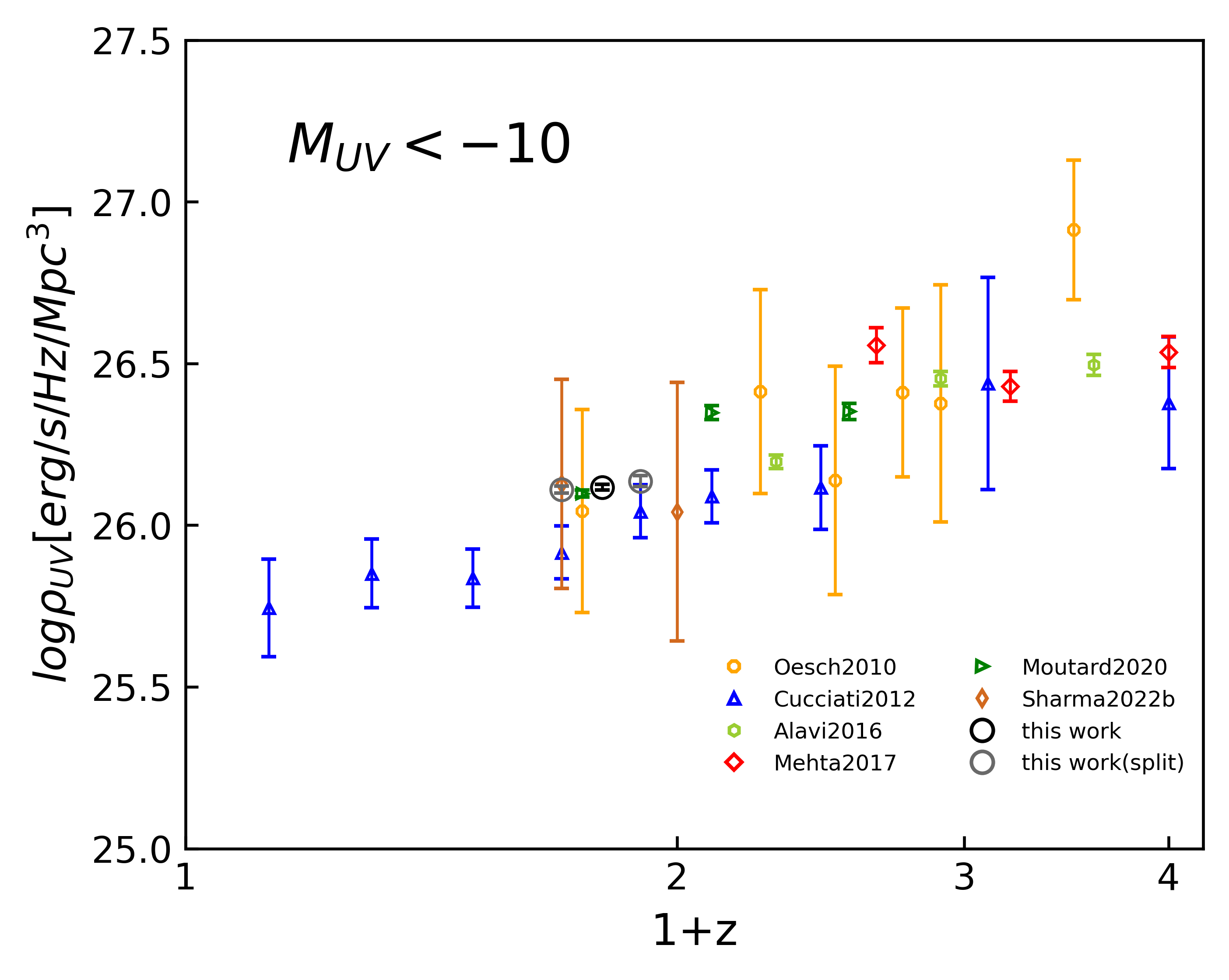}
\caption{Redshift evolution of the UV luminosity density. The black open circles shows the results computed from the best-fit Schechter paramters of our full sample at $z\sim0.8$ (not corrected for dust), and the gray circles represent the results of our split sub-samples at $z\sim0.7$ and $z\sim0.9$. Symbols representing results from literature are summarized in the legend. All the points are derived by integrating the rest-frame UV LFs down to $M_{UV}=-10$ and the $1\sigma$ errors on the points are estimated using the uncertainties in the LF parameters reported in each individual reference. }
\label{fig:rho-evol}
\end{figure}

We list in Table~\ref{tab:pars} the cumulative UV luminosity density computed by integrating down to a magnitude limit of $M_\text{UV}=-10$. We adopt this magnitude limit for convenience of comparison, since it is commonly used in the literature (e.g., \citealt{Alavi16} and \citealt{Sharma22b}). The evolution of the UV luminosity density over redshift $0<z<3$ is also illustrated in Figure~\ref{fig:rho-evol}. All results here are obtained by integrating down to $M_\text{UV}=-10$ according to Equation~\ref{eq:uvld} in a consistent manner, using the published Schechter parameters from the literature along with their uncertainties \citep{Oesch10, Cucciati12, Alavi16, mehtaUVUDFUVLuminosity2017, Moutard20, Sharma22b} . We assume there is no turnover on the UV LFs down to this absolute magnitude. To estimate the $1\sigma$ uncertainty of $\rho_\mathrm{UV}$, we run MCMC sampling to obtain a sequence of random pairs of $(\alpha,M^{*})$ using their 2D joint probability distribution and then compute the distribution of UV luminosity density and the corresponding uncertainty.
Overall, our estimated $\rho_\text{UV}$ using the total sample at $z=0.8$ is in good agreements (within $1\sigma$) with results from \citet{Oesch10} and from \citet{Moutard20}, both given at $z=0.75$. Combining all these results, the unobscured UV luminosity density continuously increases from $z=0$ to $z=2$, as also found in previous studies (e.g., \citealt{Cucciati12}; \citealt{Alavi16}). \citet{Bouwens22} show that this trend is still maintained for the dust-corrected UV luminosity density at the same redshift range. Since the integrated UV luminosity density is tightly related to the star formation rate (SFR) density, it indicates that the cosmic SFR is gradually rising from $z=0$ back to $z=2$.

\section{Summary}
\label{sec:summary}

The unprecedented deep-field UV (F275W) area coverage of UVCANDELS enables a high precision measurement of the rest-frame UV LFs. In addition, our UV-optimized aperture photometry method yields a factor of $1.5\times$ increase in the SNRs of our F275W imaging. We present the photometric catalogs of our UVCANDELS F275W and F435W image, measured using the V-isophot aperture photometry method, developed in \citet{rafelskiUvudfUltravioletNearinfrared2015}. Using well tested photometric redshift measurements and SNR cut, we identify in total 5,810 galaxies from the F275W catalog at a redshift range of $0.6 < z < 1$, down to an absolute magnitude of $M_{UV} = -14.2$. We restrict our analysis to sources above $30\%$ completeness to minimize the effect of uncertainties in estimating the completeness function, especially at the faint-end, which provides a final sample of 4726 galaxies at $-21.5<M_\text{UV}<-15.5$.

We perform a maximum likelihood estimate on the unbinned data to derive the best-fit Schechter parameters of UV LF. Overall, our best-fit Schechter parameters $\alpha$ and $M^{*}$ at $z\sim0.8$ are in good agreement with the results from \citet{Weisz14} at a similar redshift $z\sim0.75$. We are also in good agreement with \citet{Bhattacharya23} given their large uncertainties. Especially, we both infer a relatively larger $M^{*}\sim-18.5$ at $z\sim0.75-0.8$. 
We report a best-fit faint-end slope of $\alpha=-1.359^{+0.041}_{-0.041}$ at $z\sim0.8$. 

To further investigate the evolution of the UV LF, we split our full sample into 2 sub-samples at different redshift ranges. The resulting best-fit values of faint-end slope $\alpha$ are $-1.322^{+0.057}_{-0.056}$ at $z\sim0.7$ and $-1.419^{+0.067}_{-0.066}$ at $z\sim0.9$. Although the $\alpha$ values at both redshifts are consistent at a $3\sigma$ level, our results suggest that the faint end slope of galaxy UV LFs may be getting steeper with redshift. Combining our determinations with previous studies from different redshifts, this evolving trend of $\alpha$ can be more clearly observed at a larger baseline of redshift. This provides further support to the link between the build-up of galaxies and their dark matter halos. 

We obtain an unobscured cumulative UV luminosity density down to $M_\text{UV}<-10$ assuming there is no turnover on UV LFs. Our estimated $\rho_\text{UV}$ using the total sample at $z=0.8$ is in good agreement (within $1\sigma$) with results from \citet{Oesch10} and from \citet{Moutard20} both given at $z=0.75$. Taking into account that the faint end slope of UV LF is probably getting steeper with redshift, low luminosity galaxies can therefore contribute more to the total cosmic UV luminosity density at higher redshift. Combining with the Lyman-continuum escape fraction study, it is promising to illuminate the role of faint galaxies play in the cosmic reionization.

\begin{acknowledgments}
This work is based on observations with the NASA/ESA Hubble Space Telescope obtained at the Space Telescope Science Institute, which is operated by the Association of Universities for Research in Astronomy, Incorporated, under NASA contract NAS5-26555. Support for Program number HST-GO-15647 was provided through a grant from the STScI under NASA contract NAS5-26555. We are greatly appreciative to Dale Kocevski for providing us the AGN catalogs to cross-match and purify our samples. We thank the referee for the very constructive comments, which help to improve the paper greatly. We also thank Cheng Cheng and Nicha Leethochawalit for very helpful discussion on K-corrections and completeness simulations, respectively. XW is supported by the National Natural Science Foundation of China (grant 12373009), the CAS Project for Young Scientists in Basic Research Grant No. YSBR-062, the Fundamental Research Funds for the Central Universities, and the Xiaomi Young Talents Program.
\end{acknowledgments}


\end{document}